\documentclass[apj]{emulateapj}

  \usepackage{natbib}
  \bibpunct{(}{)}{;}{a}{}{,} 

\newcommand{\mz}{{\small $\mathcal{M}$-Z}}
\newcommand{\Sz}{{\small {\large $\Sigma$}-Z}}

\DeclareRobustCommand{\ion}[2]{%
\relax\ifmmode
\ifx\testbx\f@series
{\mathbf{#1\,\mathsc{#2}}}\else
{\mathrm{#1\,\mathsc{#2}}}\fi
\else\textup{#1\,{\mdseries\textsc{#2}}}%
\fi}
\newcommand{\HII}{\ion{H}{ii}~}

\shorttitle{The CALIFA survey}
\shortauthors{S\'anchez et al.}

\begin{document}

\title{The CALIFA Survey: A Panoramic view on Galaxy properties}

\author{S.F~S\'anchez\altaffilmark{1}}

\altaffiltext{1}{Instituto de Astronom\'\i a,Universidad Nacional Auton\'oma de Mexico, A.P. 70-264, 04510, M\'exico,D.F.}

\email{sfsanchez@astro.unam.mx}

\begin{abstract}

We present here a brief summary of the status of the on-going CALIFA
survey. We have just started the last semester of observing (Spring
2015). So far, we have gathered IFU data of more than 600 galaxies,
$\sim$85\% of them corresponding to the main CALIFA sample (516
objects). We give an overview of some of the main science results that
have been published by the CALIFA team during the last four years. In
particular, we emphasise the results regarding the properties of the
ionized gas in galaxies and the gradients in oxygen abundance, as well
as the evidence for inside-out growth of galaxies uncovered through
analysis of the stellar population content.

\end{abstract}



\section{Introduction \label{intro}}

The Calar Alto Legacy Integral Field Area (CALIFA) survey
\citep{sanchez12} is an ongoing large project of the Centro
Astron\'omico Hispano-Alem\'an at the Calar Alto observatory to obtain
spatially resolved spectra for 600 local (0.005$<z<$0.03) galaxies by
means of integral field spectroscopy (IFS). CALIFA observations
started in June 2010 with the Potsdam Multi Aperture Spectrograph
\citep[PMAS,][]{roth05}, mounted to the 3.5 m telescope, utilizing the large
(74$\arcsec$$\times$64$\arcsec$) hexagonal field-of-view (FoV) offered by the PPak
fiber bundle \citep{verheijen04,kelz06}. PPak was created for the Disk
Mass Survey (Bershady et al. 2010). Each galaxy is observed using two
different setups, an intermediate spectral resolution one (V1200,
$R\sim 1650$), that covers the blue range of the optical wavelength
range (3700-4700\AA), and a low-resolution one (V500, $R\sim 850$,
that covers the first octave of the optical wavelength range
(3750-7500\AA). A dithering technique was applied to guarantee the
full coverage of FoV, providing with 993 independent spectra of each
galaxy, and with a final cube of $\sim$2500 spectra of
1$\arcsec$/spaxel sampling and $\sim$2.5$\arcsec$ FWHM of spatial
resolution (per setup).

A diameter-selected sample of 939 galaxies was drawn from the 7th data
release of the Sloan Digital Sky Survey (SDSS), described in
\cite{walcher14}. In summary, we selected galaxies within a certain
redshift range ($0.005 < z < 0.03$), and optical extension that fits
within the FoV of the instrument ($45'' < D_{25} < 80''$, where
$D_{25}$ is the isophotal diameter in the SDSS $r$-band).  The CALIFA
main observed sample is a sub-selection from this mother sample, based
on visibility only. We have currently observed 516 of these objects
(December 2014). In addition, several projects have targeted a
heterogenous sample of more than 100 galaxies that comprise the CALIFA
extended sample, including (i) objects that are in low numbers in the
main sample: dwarf galaxies, faint and compact E/S0, compact blue
elliptical galaxies; and (ii) companions in merging/interacting
systems when the main galaxy was selected within the CALIFA sample
criteria. This extended sample does not fulfill all the selection
criteria of the main sample, however, it comprises galaxies that are
mostly in the same foot-print in terms of redshift and diameter (the
main selection criteria). Both the main and extended samples were
observed using the same instrumental configuration, and reduced using
the same pipeline, that guarantees the homogeneity of the
dataset. Alltogether, we have gathered IFU data for more than 600
galaxies.

\begin{figure*}
\resizebox{\hsize}{!}
{\includegraphics[width=7.6cm]{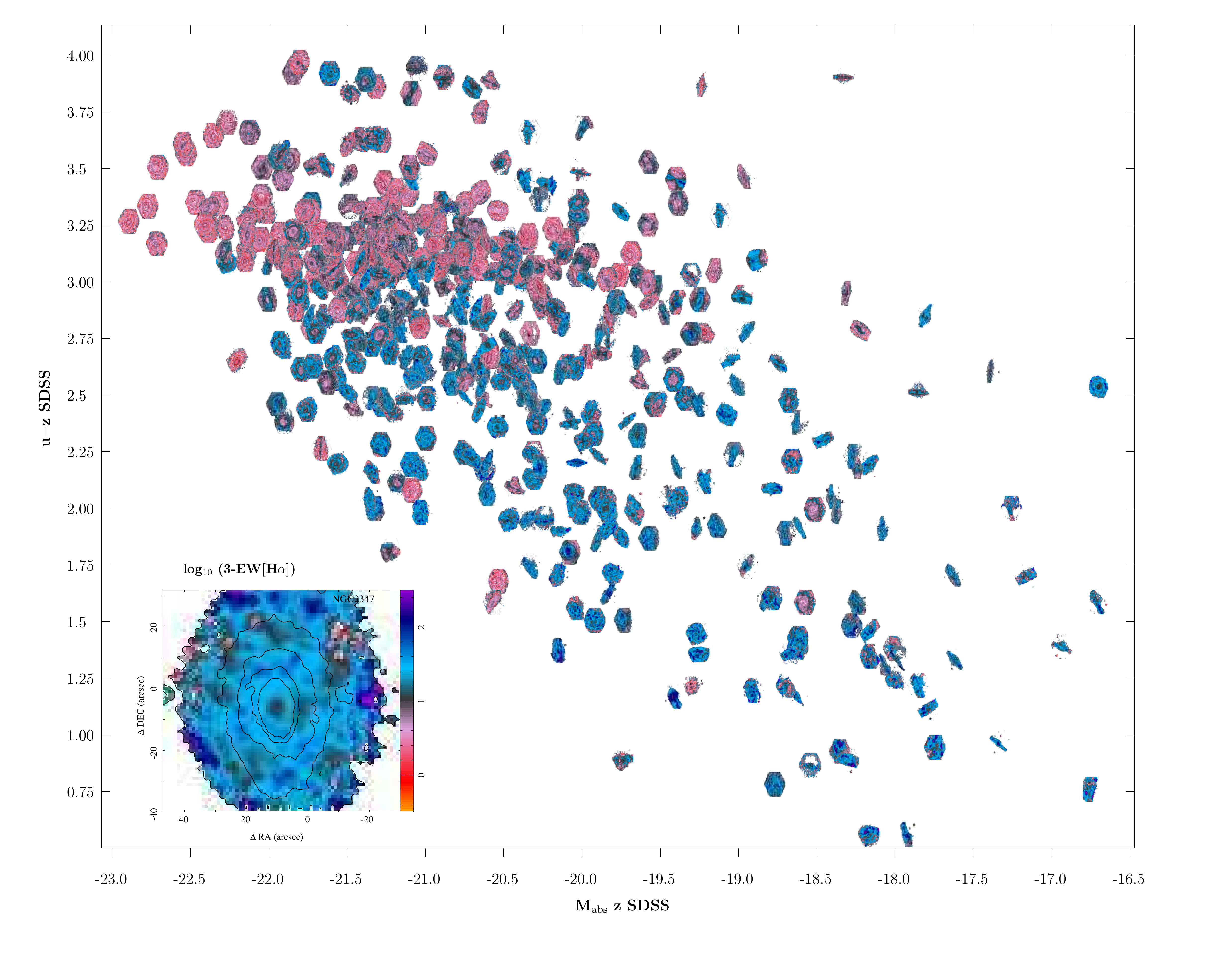}
\includegraphics[width=7.6cm]{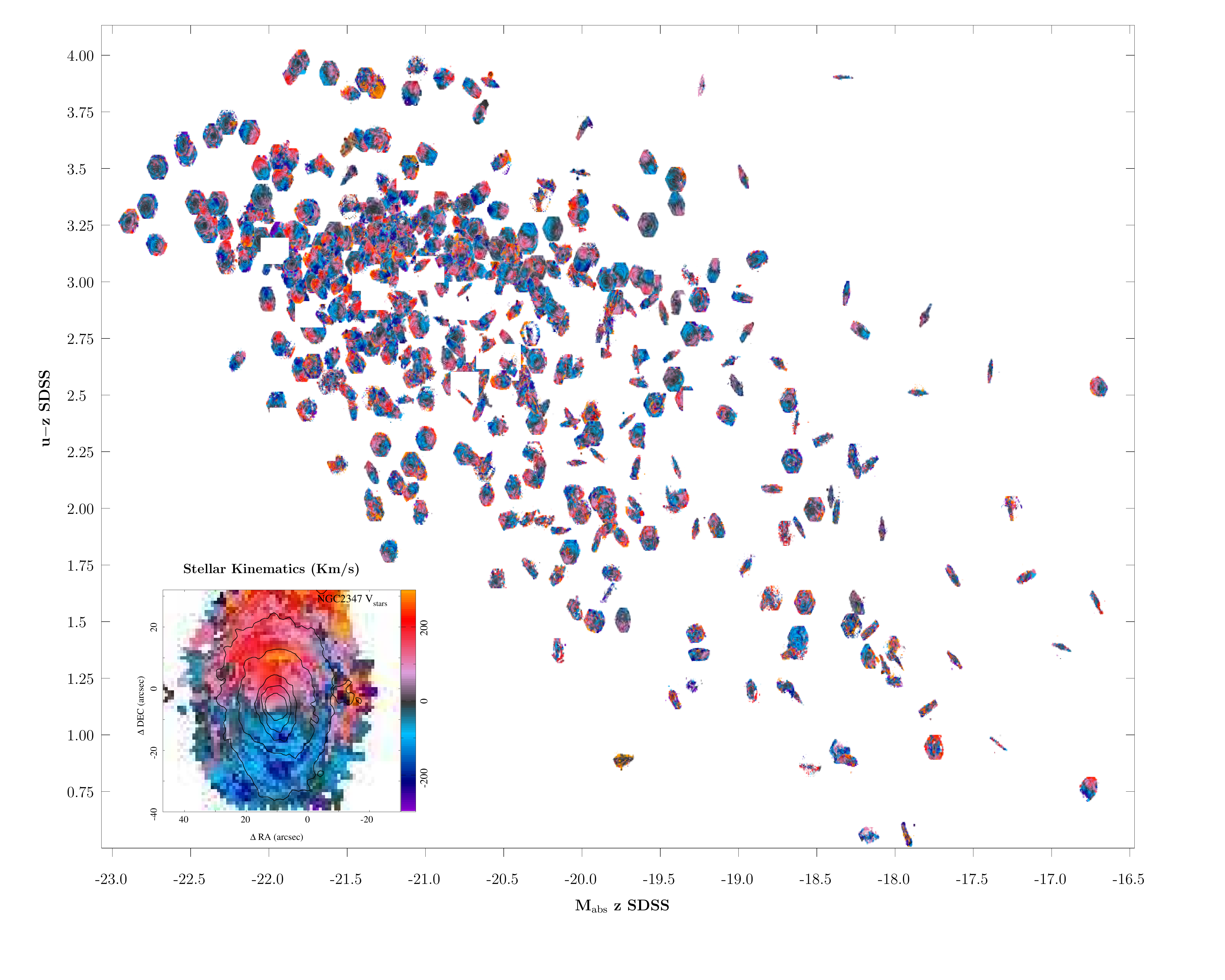}}
  \caption{\label{data} 
\label{fig:MZ}. {\it Left panel:} Maps of the equivalent width of H$\alpha$ for the 516 galaxies of the CALIFA main sample
observed up to December 2014 distributed along the Color-Magnitude diagram. The blue/dark-blue areas in each map correspond to the areas of intense star-formation (if distributed across the galaxy) or affected by ionization due to an AGN (if concentrated in the center). The pink areas corresopond to retired areas (or retired galaxies), i.e., areas where the starformation has stopped, and they are most probably ionized by post-AGBs. {\it Right Panel:} Similar distribution for the stellar velocity maps of the same galaxies. The blue areas correspond to approaching movements, while the red areas correpond to receeding movements. In both panels the box in the bottom-left shows an enlarged version of the corresponding maps for the galaxy NGCº,
 }
\end{figure*}

Combining the techniques of imaging and spectroscopy through optical
IFS provides a more comprehensive view of individual galaxy properties
than any traditional survey. CALIFA-like observations were collected
during the feasibility studies (M\'armol-Queralt\'o et al. 2011; Viironen
et al. 2012) and the PPak IFS Nearby Galaxy Survey (PINGS,
Rosales-Ortega et al. 2010), a predecessor of this survey. First
results based on those datasets already explored their information
content (e.g. Rosales-Ortega et al. 2010;  Rosales-Ortega et
al. 2012). 

Compared with other IFS surveys, CALIFA offers a unique combination of
(i) a sample covering a wide range of morphological types in a wide
range of masses, sampling the Color-Magnitude diagram for M$_g>-$ 18
mag; (ii) a large FoV, that guarantees to cover the entire optical
extent of the galaxies up to 2.5$r_e$ for an 80\% of the sample; and
(iii) an accurate spatial sampling, with a typical spatial resolution
of $\sim$1 kpc for the entire sample, which allows to obtain spatially
resolved spectroscopic properties of most relevant structures in
galaxies (spiral arms, bars, bulges, \ion{H}{ii} regions...). The
penalty for a better spatial sampling of the galaxies is the somewhat
limited number of galaxies in the survey, compared to more recently
started ones, e.g., MaNGA \citep[$\sim$ 10,000 galaxies, ][]{bundy14}
and SAMI \citep[$\sim$3,600 galaxies,][]{sami}, or the foreseen ones,
e.g., HECTOR \citep[$\sim$100,000 galaxies,][]{joss14}.  Like in the
case of the CALIFA extended sample ($\sim$15\% of the total objects),
not all the MaNGA and SAMI targets have been selected using a single
criteria. In MaNGA there are two main selection criteria
that comprises galaxies to be covered up to $\sim$1.5$r_e$ ($\sim$70\%
of the sample), and up to $\sim$2.5$r_e$ ($\sim$20\% of the sample),
both of them at a different average redshift, and finally there is a
set of so-called ancillary projects that will comprise 5-10\% of the
total observed objects \citep{bundy14}.  In SAMI the main sample was
built-up using a set of volume limited samples at different redshift
ranges ($\sim$70\% of the sample), with an additional sample built-up
by galaxies in clusters \citep{sami15}. In terms of the spectral
resolution, while in the red both SAMI and MaNGA surveys have better
spectral resolutions than CALIFA (in particular SAMI), in the blue
the three have similar resolutions.

As a legacy survey, one of the main goals of the CALIFA collaboration
is to grant public access of the fully reduced datacubes. In November
2012 we delivered our 1st Data Release (Husemann et al. 2013),
comprising 200 datacubes corresponding to 100
objects. After almost two years,
and a major improvement in the data reduction, we present our 2nd Data
Release (Garc\'\i a Benito et al., 2014), comprising 400 datacubes
corresponding to 200 objects\footnote{http://califa.caha.es/DR2/},
the 1st of October 2014.  The final Data Release, comprising the full
dataset is foreseen for spring 2016. This DR will present
the data using an updated version of the reduction pipeline, version 2.0, 
that is currently under development.

\section{CALIFA: Main Science Results}

The data products that can be derived from the IFU datasets obtained
by the CALIFA survey comprise information on the stellar populations,
ionized gas, mass distribution and stellar and gas kinematics. Figure
\ref{data} shows an example of two of the indicated dataproducts for
the 516 galaxies of the main sample observed up to December 2014: (i)
the equivalent width of H$\alpha$, that traces the areas of
star-formation across the galaxies, being proportional to the specific
star-formation rate and (ii) the stellar velocity maps obtained using
the low-resolution V500 setup (better quality ones are derived using
the high-resolution V1200). These dataproducts are presented along the
Color-Magnitude diagram. The red-sequence of early-type/dry galaxies
is clearly identified as a pink-sequence in the left-hand panel,
illustrating the location of those galaxies with little or none
star-formation. In some cases there are clear blue-spots in the center
of those galaxies, indicating, most likely, the presence of an AGN.
The blue cloud is dominated by galaxies with evident star-formation
all over its optical extension, with the spiral arms clearly
identified as areas of dark-blue color, corresponding to \ion{H}{ii}
regions or agregations.

The bimodal distribution clearly visible in the
distribution of EW(H$\alpha$) along the color-magnitud diagram is
less evident in the distribution of stellar velocity maps. While most
of the slow-rotating systems are located along the red-sequence, there
is a considerable number of fast-rotating objects at a similar
location. Only the combination of this information with the
distribution of stellar velocity dispersion clarifies if those
galaxies are pressure or rotationaly supported. In summary, both
panels illustrate how CALIFA conforms a panoramic view of the spatial
resolved spectroscopic properties of galaxies in the Local Universe.

\begin{figure*}
\resizebox{\hsize}{!}
{\includegraphics[width=7.6cm]{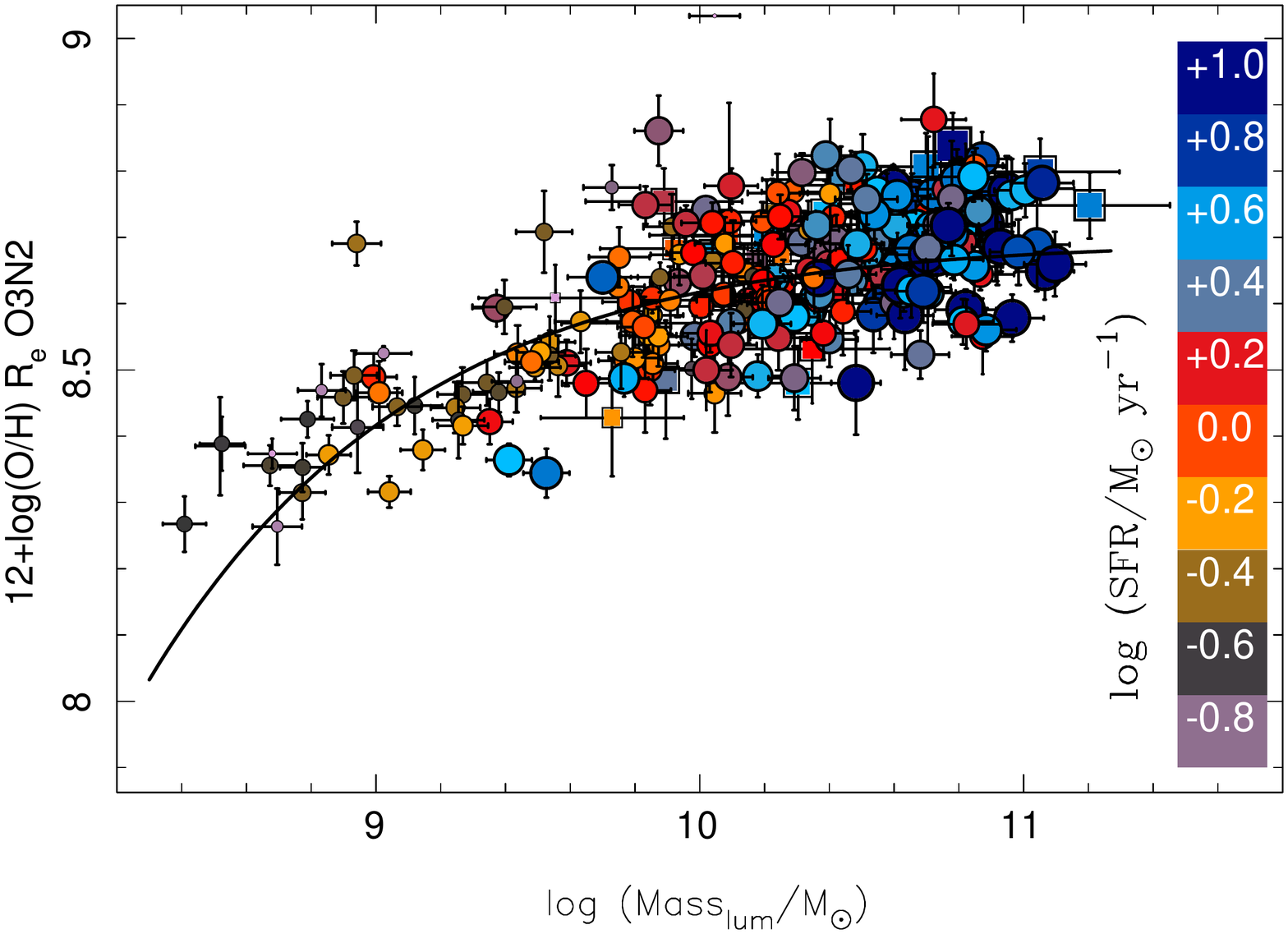}
\includegraphics[width=7.6cm]{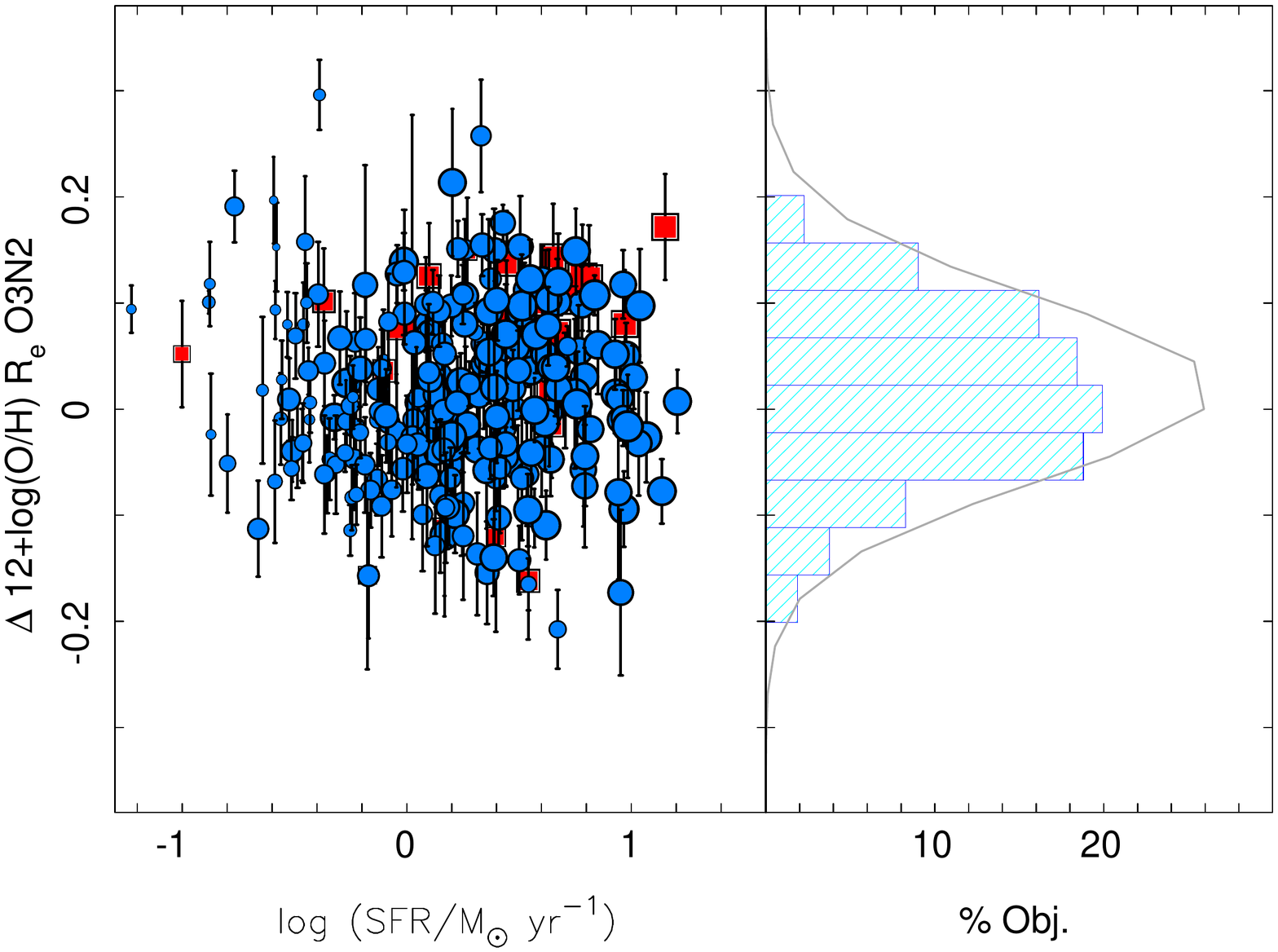}}
  \caption{\label{spec} 
\label{fig:MZ}. {\it Left panel:} Distribution of the oxygen abundances at the effective radii as a function of the integrated stellar masses for the CALIFA galaxies (236, circles), together with those from the CALIFA feasibility studies (31, squares). {\it Right Panel:} Distribution of the differential oxygen abundances with respect to the solid-line shown in the left-panel (i.e., the dependence on the stellar mass), as a function of the integrated SFR for the CALIFA galaxies.
 }
\end{figure*}

Different science goals have been already addressed using this
information: (i) New techniques have been developed to understand the
spatially resolved star formation histories (SFH) of galaxies (Cid
Fernandes et al., 2013, 2014). We found the solid evidence that
mass-assembly in the typical galaxies happens from inside-out \citep{perez13}. The SFH and chemical enrichment of bulges and early-type galaxies are
fundamentally related to the total stellar mass, while for disk
galaxies it is more related to the local stellar mass density
\citep{rosa13,rosa14}; negative age gradients indicate that 
quenching progresses outward in massive galaxies \citep{rosa13},
and age and metallicity gradients  of spirals are not altered significantly by
the presence of a bar \citep{patri14}; (ii) We explore the origin
of the low intensity, LINER-like, ionized gas in galaxies. These
regions are clearly not related to star-formation activity, or
to AGN activity. They are most probably related to post-AGB ionization
in many cases \citep{papa13}; (iii) We study the effects of the aperture and resolution on IFS data. CALIFA provides a unique tool to understand the aperture and
resolution effects in larger single-fiber (like SDSS) and IFS surveys
(like MaNGA, SAMI). We explored the effects of the dilution of the
signal in different gas and stellar population properties (Mast et
al., 2014), and proposed an new empirical aperture correction for the
SDSS data \citep{iglesias13}; (iv) CALIFA is the first
IFU survey that allows gas and stellar kinematic studies for all
morphologies with enough spectroscopic resolution to study (a) the
kinematics of the ionized gas \citep{bego14}, (b) the
effects of bars in the kinematics of galaxies \citep{jkbb14}; (c) the effects of the interaction stage on the kinematic
signatures (Barrera-Ballesteros et al., submitted), (d) measure the
Bar Pattern Speeds in late-type galaxies (Aguerri et al., accepted),
(v) extend the measurements of the angular momentum of galaxies to
previously unexplored ranges of morphology and ellipticity
(Falc\'on-Barroso et al., in prep.); (vi) using CALIFA data we probed for the first time spectroscopically the different association of different SN types to the star-formation of their environment \citep{lluis14}; and (vii) finally we analyze in
detail the effects of galaxy interaction in the enhancement of
star-formation rate and the ignition of galactic outflows \citep{wild14}. 

\section{Results of our studies of the \HII regions}\label{high}

The program to derive the properties of the \HII regions
was initiated based on the data from the PINGS survey
\citet{rosales-ortega10}. This survey acquired IFS mosaic data for a dozen
of medium size nearby galaxies. In \citep{sanchez11} and
\citet{rosales11} we studied in detail the ionized gas and \ion{H}{ii}
regions of the largest galaxy in the sample (NGC\,628). The main
results of these studies are included in the contribution by
Rosales-Ortega in the current edition. We then continued the
acquisition of IFS data for a larger sample of visually classified
face-on spiral galaxies \citep{marmol-queralto11}, as part of the
feasibility studies for the CALIFA survey \citep{sanchez12}. The
spatially resolved properties of a typical galaxy in this sample,
UGC9837, were presented by \cite{viir12}.

In \cite{sanchez12b} we presented a new method to detect, segregate
and extract the main spectroscopic properties of \ion{H}{ii} regions
from IFS data
(\textsc{HIIexplorer}). A preliminar catalog of  $\sim$2600 \ion{H}{ii} regions and aggregations
extracted from 38 face-on spiral galaxies compiled from the PINGS and
CALIFA feasibility studies was presented. We found
a new local scaling relation between the stellar mass density and
oxygen abundance, the so-called \Sz\ relation \citep{rosales12}.

The same catalog allows us to explore the galactocentric radial gradient
of the oxygen abundance \citep{sanchez12b}. We confirmed that up to
$\sim$2 disk effective radius there is a negative gradient of the
oxygen abundance in all the analyzed spiral galaxies. The gradient
presents a very similar slope for all the galaxies ($\sim -0.12$
dex/$r_e$), when the radial distances are measured in units of the
disk effective radii. Beyond $\sim$2 disk effective radii our data show evidence of a
flattening in the abundance, consistent with several other
spectroscopic explorations, based mostly on a few objects
\citep[e.g.][]{bresolin09}.

In \citep{sanchez13} we presented the first results based on the
catalog of \HII\ regions extracted from a enlarged sample of
galaxies ($\sim$100). We studied the dependence of the \mz\ relation
with the star formation rate. We found
no secondary relation different from the one induced by the well
known relation between the star formation and the mass, contrary to
what was claimed other authors \citep{lara10a,mann10}, based on single
aperture spectroscopic data (SDSS). Although the reason for the
discrepancy is still not clear, we postulate that simple aperture
bias, like the one present in previous datasets, may induce the
reported secondary relation. Figure \ref{fig:MZ} presents an updated
version of these results, including the last list of analyzed
galaxies, until July 2014 (236 galaxies from the CALIFA sample
together with 31 galaxies from the CALIFA-pilot studies). The left
panel shows the \mz\ relation found for these galaxies, with color
code indicating the integrated SFR for each galaxy. It is appreciated
that the stronger gradient in SFR is along the stellar mass, as
expected for star-forming galaxies. Once subtracted the best fitted
function to the \mz\ relation, the residuals of the abundance do not
present any evident secondary relation with the SFR (as it is seen in
the right panel). Thus, the results presented in \citep{sanchez13} are
confirmed with a sample of galaxies enlarged by almost a factor two.

We also confirmed the local \Sz\ relation unveiled by
\citet{rosales11}, with a larger statistical sample of \HII regions
($\sim$5000). This nebular gas \Sz\ relation is flatter than the one
derived for the average stellar populations \citep{rosa14}, but both
of them agree for the younger stars, as expected if the most recent
stars are born from the chemically enriched ISM.  In \cite{sanchez14}, we
confirmed that the abundance gradients present a common slope up to
$\sim$2 effective radii, with a distribution compatible with being
produced by random fluctuations, for all galaxies when normalized to
the disk effective radius of $\alpha_{O/H}=-$0.1 dex/$r_e$.  Similar
results are found when the gradient is normalized to other
scale-lengths of the galaxy, like $r_{25}$, with a sharper slope,
$\alpha_{O/H} = -$0.16 dex/$r_{25}$ \citep{sanchez14}.  When using the
physical galactocentric distance the gradient is shallower
($\alpha_{O/H} = -$0.03 dex/kpc) and presents a tail towards large
slopes, up to $-$0.15 dex/kpc, with a clear morphological
dependence. The use of different abundance calibrators does not affect
the main conclusion, although it changes the numerical value of the
gradient, as recently published by \cite{ho15}, using CALIFA-DR1 and
CALIFA-feasibility study data.

Finally, in \cite{sanchez14b}, we found evidence that \HII regions
keep a memory of their past, by analysing the correspondance between
the properties of these ionized regiones with that of their underlying
stellar populations. 

\section{Conclusions}

In summary the results from the CALIFA survey present a coherent picture of
the mass-growth and chemical enrichment of galaxies. All the results
indicate, so far, that the bulk of the galaxy population presents an
inside-out growth (at the mass range covered by the survey),
with a chemical enrichment dominated by local processes, and
limited effects by processes like outflows or radial mixing.

We are still analysing the data, and in particular we are trying
to uncover the chemical enrichment processes at different ages using
inverse methods to derive them form the stellar populations (Gonzalez Delgado 
et al., in prep.), and the current picture could be modified on the
basis of the new results.

\section{Acknowledgments}

CALIFA is a collaborative effort, and the summarized results are the
scientific products of all its members. I cannot enlist all of them in
this proceedings, but I thank them for their outstanding job. I also
thank the Calar Alto Observatory and its staff for the great
involvement and support to this project. I thank the support of the
ConaCyt-180125-22079-132 and PAPIIT-IA100815 grants.


 \label{lastpage}


\end{document}